\documentclass[preprint,showpacs,aps,preprintnumbers,amsmath,amssymb,
nofootinbib]{revtex4}
\usepackage{epsfig}
\begin{document}

\begin{flushright}
\end{flushright}


\newcommand{\be}{\begin{equation}}
\newcommand{\ee}{\end{equation}}
\newcommand{\bea}{\begin{eqnarray}}
\newcommand{\eea}{\end{eqnarray}}
\newcommand{\bers}{\begin{eqnarray*}}
\newcommand{\eers}{\end{eqnarray*}}
\newcommand{\nn}{\nonumber}

\title{\large Determining the CKM angle $\gamma$ with $B_c$ decays  }
\author{A. K. Giri$^1$, B. Mawlong$^2$ and R. Mohanta$^2$ }
\affiliation{$^1$Physics Department, Punjabi University,
Patiala-147002, India\\
$^2$School of Physics, University of Hyderabad, Hyderabad-500 046, India\\}

\vspace*{0.2 truein}
\begin{abstract}
We consider the possibility of extracting the CKM angle $\gamma$ with
$B_c$ decays. The modes $B_c^\pm \to (D^0) D_s^\pm \to
(K^{*+}K^-) D_s^\pm $ and $B_c^\pm \to  (\bar D^0) D_s^\pm \to
(K^{*+}K^-) D_s^\pm $ are found to be well suited for
the extraction of $\gamma$.  Since a
large number of $B_c$ mesons are expected to be produced at the
LHC, it would be very interesting to explore the determination of $\gamma $
with these modes.
\end{abstract}
\pacs{12.15.Hh, 13.25.Hw, 11.30.Er }
\maketitle

It is strongly believed that the elusive Higgs boson, the missing
entity in the otherwise immensely successful standard model (SM) of
electroweak interactions, will be chased and most likely
to be found at the Large
Hadron Collider (LHC), which is going to be started very soon.
While a detailed understanding of the SM description might be
accomplished during the LHC era, there is an unprecedented level of
enthusiasm to decipher the signal of high scale physics,
where the SM is a low energy manifestation
of the same. Whether the physics at a higher scale leaves
its trace at LHC or not but it is certain that the enormous data will
provide us unique opportunity to study
all the important aspects of physics under the framework of the SM with a
greater accuracy.

In the SM, the CP violation is
elegantly described by the Cabibbo-Kobayashi-Maskawa (CKM) mechanism.
In this context, one of the main ingredients of the SM description of
CP violation is the
CKM unitarity triangle (UT) and the angles of the UT are
termed as $\alpha$ ($\phi_2$), $\beta$ ($\phi_1$) and $\gamma$ ($\phi_3$)
\cite{bdecays}. Large CP violation, as was expected, has already
been established in $B$-systems
in the currently running $B$-factories at SLAC and KEK.
The present status is that we have measured, with the
huge data sets available, the angle $\beta$ (actually, sin (2$\beta$))
with a reasonable accuracy
and we expect to have a precision measurement of angle $\beta$ in the
years to come, with the help of the golden mode $B_d^0 \to J/\psi K_S$.
Unfortunately,
we do not have three golden modes
to determine the three
angles of the UT. So we have to be contented with the best available
modes like $B\to
\pi\pi$ (and some related modes) for the determination of the angle
$\alpha$, but these modes are accompanied by a generic problem called penguin
contamination, whose remedy has not been found yet by the theoretical
community. So
finally, we are left with the angle $\gamma$ = arg$(-{V_{ud}V_{ub}^*}/
V_{cd}V_{cb}^*)$, which was believed to be the most
difficult one, among all the three angles, at the beginning.
But, fortunately, in this case, nature has been very
kind to provide us many options to determine the
angle $\gamma$ in various avenues.

There have been many attempts in the past to devise methods to determine
the CKM angle $\gamma$ as cleanly as possible.
The golden method to determine $\gamma$ is the Gronau-London-Wyler (GLW)
method \cite{glw},
which uses the interference of two amplitudes ($b\to c \bar u s$ and $b\to
u \bar c s$) in $B\to DK$ modes. In this method $\gamma$ can be 
determined by measuring the decay rates $ B^- \to D^0 K^-$, $B^- 
\to \bar D^0 K^-$ and $B^- \to D_+^0 K^-$ ( where $D_+^0$ is the CP-even 
eigenstate of neutral $D$ meson system) and their corresponding 
CP conjugate modes. However, because the mode $B^- \to \bar D^0 K^-$
is both color and CKM suppressed with respect to $B^- \to D^0 K^-$
the corresponding amplitude triangles are expected to be highly squashed and 
it is also a very difficult to measure the rate of $B^- \to \bar D^0 
K^-$. To overcome the problems of GLW method Atwood-Dunietz-Soni (ADS)
\cite{ads} proposed an improved method where they have considered the decay
chains $B^- \to K^- D^0[\to f]$ and  $B^- \to K^- \bar D^0[\to f]$,
where $f$ is the doubly Cabibbo suppressed (Cabibbo favored) 
non-CP eigenstate of $D^0(\bar D^0)$. 
These methods are being explored in the currently
running $B$-factory experiments and will also be taken up at
the collider experiments alongwith another golden
method called Aleksan-Dunietz-Kayser (ADK) method \cite{adk}, which uses the
time dependent measurement of $B_s^0 (\bar B_s^0)\to D_s^\mp K^\pm$ modes.
Because of its importance and, of course,
possible options available there are many methods that exist in the
literature. Some of the alternative methods to obtain
$\gamma$ are those using $B$ and $B_s$ decays \cite{ggsz}-\cite{ns},
 $B_c$ decays \cite{bc} and also  $\Lambda_b$ decays \cite{gkm2}.

In the meantime, another exciting method,
Giri-Grossman-Soffer-Zupan (GGSZ) method (otherwise also known as
the Dalitz method) \cite{ggsz}
has been proposed (using $B\to D^0(\bar D^0) K \to   K_S \pi \pi K $),
which has many attractive features and has already been explored at both the
$B$-factories. It
should be noted here that the GGSZ method uses the ingredients of GLW and
ADS method where the $D^0 (\bar D^0)$ decays to multi-particle final states.
This method in turn helps us to constrain the angle $\gamma$
directly from the experiments. But at present the error bars
are quite large, which are expected to come down in the coming years. It may
be worthwhile to emphasize here
that one has to measure the angle with all possible clean methods available
to arrive at a conclusion and thereby reducing the error in $\gamma$ to
a minimum.

In this continued effort, we now wish to explore yet another method
with the decays
$B_c^\pm \to D_s^\pm D^0\to D_s^\pm (K^{*+}K^-)_{D^0}$
and  $B_c^\pm \to D_s^\pm
\bar D^0\to D_s^\pm (K^{*+}K^-)_{\bar D^0}$. It has been shown
earlier in \cite{bc}
 that the decay $B_c^\pm\to D^0(\bar D^0)D_s^\pm $ modes
can be used to determine the CKM angle $\gamma$ in a better way
since the interfering amplitudes in $B_c$ case are 
roughly of equal sizes, whereas the
corresponding ones in GLW method (using $B$ mesons) are not so.
In our earlier work \cite{bc}, we have shown that $\gamma$ can be 
determined from the decay rates $B_c^\pm \to D^0 D_s^\pm$,  
$B_c^\pm \to \bar D^0 D_s^\pm$ and  $B_c^\pm \to D_\pm^0 D_s^\pm$
(where $D_\pm^0$ are the CP eigenstates of neutral $D$
meson system with CP eigenvalues $\pm 1$, which can be identified
by the CP-even and CP-odd decay products of neutral $D$ meson).
In this work we propose another method where
we consider the $B_c^\pm \to D^0 (\bar D^0)D_s^\pm$
decay modes, that are followed by  $D^0(\bar D^0)$
decaying to $K^{*+}K^-$, which is a non-CP eigenstate.

The decay modes $B_c^- \to D_s^- D^0 $ and $B_c^- \to D_s^- \bar D^0$
are described by the quark level transition $b \to c \bar u s $ and $b \to
u \bar c s$ respectively and the amplitudes for these  processes are given as
\bea
{\cal A}(B_c^- \to D^0 D_s^-) =  \frac{G_F}{\sqrt{2}} V_{cb}V_{us}^*
(C+A)\;,~~~
{\cal A}(B_c^- \to \bar D^0 D_s^-) =  \frac{G_F}{\sqrt{2}} V_{ub}V_{cs}^*
(\tilde C+ \tilde T)\;,
\eea
where $C$ and $A$ denote the color suppressed tree and annihilation topologies
for $b \to c$ transition and $\tilde C$ and $\tilde T$ denote the color
suppressed tree and color allowed tree contributions for $b \to u$ transition.
It should be noted here that the amplitude with the smaller CKM element
$V_{ub}$ is color allowed while the larger element $V_{cb}$ comes
with color suppression factor (and alongwith the the appropriate
$ V_{cs}$ and $V_{us}$ elements) the two amplitudes are of comparable
sizes.
Now let us denote these amplitudes as
\be
A_B= {\cal A}(B_c^- \to D^0 D_s^-)\;,~~~~\bar A_B={\cal A}(B_c^- \to
\bar D^0 D_s^-)\;,
\ee
and their ratios as
\be
\frac{\bar A_B}{A_B}=r_B e^{i(\delta_B -\gamma)}\;,
~~~{\rm with}~~~r_B=\left |\frac{\bar A_B}{A_B} \right |~~~
{\rm and}~~~ {\rm arg} \left (\bar A_B/A_B \right )=\delta_B-\gamma\;,
\ee
where $\delta_B$ and $(-\gamma)$ are the relative strong and weak 
phases between the two amplitudes.
The ratio of the corresponding CP conjugate processes are obtained by
changing the sign of the weak phase $\gamma$.
One can then obtain a rough estimate of $r_B$ from dimensional analysis, i.e.,
\be
r_B=\left |\frac{V_{ub} V_{cs}^*}{V_{cb} V_{us}^*} \right |
\cdot \frac{a_1^{eff}}{
a_2^{eff}} \approx {\cal O}(1)\;,
\ee
where $a_{1}^{eff}$ and $a_2^{eff}$ are the effective QCD
coefficients describing the color allowed and color suppressed
tree level transitions. For the sake of comparison, we would like to
point out here that the 
corresponding ratio between the $B^- \to D^0(\bar D^0)K^- $ 
amplitudes are given as $|{\cal A}(B^- \to \bar D^0 K^-)/{\cal A}(B^- \to 
D^0 K^-)|=|(V_{ub} V_{cs}^*)/(V_{cb} V_{us}^*)|
\cdot (a_2^{eff}/
a_1^{eff}) \approx {\cal O}(0.1)$. 
The $D^0$ decay amplitudes are denoted as
\be
A_D= {\cal A}(D^0 \to K^{*+} K^-)\;,~~~~\bar A_D={\cal A}(
\bar D^0 \to
K^{*+} K^-)\;,
\ee
and their ratios as
\be
\frac{\bar A_D}{A_D}=r_D e^{i\delta_D}\;,
~~~{\rm with}~~~r_D=\left |\frac{\bar A_D}{A_D} \right |\;.
\ee
It is interesting to note that the parameters $r_D$ and $\delta_D$
have recently been measured by CLEO collaboration
\cite{cleo}, with values $r_D=0.52\pm0.05\pm0.04$ and $\delta_D=332^\circ
\pm 8^\circ \pm 11^\circ $, rendering our study, at this point of time,
more appealing.

With these definitions the four amplitudes are given as
\bea
{\cal A}(B_c^- \to  D_s^- (K^{*+} K^-)_D)
&=& |A_B A_D|\Big[1+r_B r_D e^{i(\delta_B+
\delta_D-\gamma)}\Big]\;,\nn\\
{\cal A}(B_c^- \to  D_s^- (K^{*-} K^+)_D)
&=& |A_B A_D| e^{i \delta_D}\Big[r_D+ r_B e^{i(\delta_B-
\delta_D-\gamma)}\Big]\;,\nn\\
{\cal A}(B_c^+ \to  D_s^+ (K^{*-} K^+)_D)
&=& |A_B A_D|\Big[1+r_B r_D e^{i(\delta_B+
\delta_D+\gamma)}\Big]\;,\nn\\
{\cal A}(B_c^+ \to  D_s^+ (K^{*+} K^-)_D) &=& |A_B A_D|e^{i
\delta_D}\Big[r_D+ r_B e^{i(\delta_B-
\delta_D+\gamma)}\Big]\;.\label{eq} 
\eea 
From these amplitudes one
can obtain the four observables ($R_1, \cdots, R_4$), with the definition
\be R_i = \left |{\cal A}_i(B_c^\mp \to  D_s^\mp (K^{*\pm}
K^\mp)_D)/{A_B A_D}\right |^2\;.\label{eq8} \ee We can now write $R_1
= 1+r_B^2 r_D^2 + 2 r_B r_D \cos (\delta_B+\delta_D -\gamma)$ and
similarly $R_2$, $R_3$ and $R_4$.

Here we assume that the amplitudes $|A_B|$ and $|A_D|$ are known
(so also $r_B$, which is ${\cal O}$(1)).

Thus, one can obtain an analytical expression for $\gamma$ as
\be
\sin^2 \gamma=\frac{ [R_1-R_3]^2
-[R_2-R_4]^2 }{4 \Big [ [R_2-(r_B^2+r_D^2)][R_4-(r_B^2+r_D^2)]
-[R_1 -(1+r_B^2 r_D^2)]
[R_3-(1+r_B^2 r_D^2)]
\Big]}\;.\label{eq1}
\ee

Now let us study the sensitivity of $\gamma $ in some
limiting cases in the method described above.\\
(a) If the relative strong phase between $\bar A_B$ and $A_B$ is zero
then Eqn. (\ref{eq1})  can no longer be used to extract the
angle $\gamma$ as both numerator and denominator vanish in
this limit. However, still $\gamma $ can be extracted, in this limit,
from either the observables $R_1$ and $R_3$ or $R_2$ and $R_4$.
Now, considering the observables $R_2$ and $R_4$, for example,
one can obtain an expression for $\gamma $ as
\be
\tan \gamma= \frac{\cot \delta_D (R_4-R_2)}{R_2+R_4-2(r_B^2+r_D^2)}\;.
\ee
Analogous expression for $\gamma $ can also be obtained from $R_1$ and $R_3$
with the replacement of $ R_{2,4} \leftrightarrow R_{3,1}$ and
$(r_B^2+r_D^2)\leftrightarrow (1+r_B^2r_D^2)$. Let us now consider another
limiting case.

(b) If $r_B=1$ and $\delta_B= 0$, then the four observables 
($R_1, \cdots, R_4$)
are no longer independent of each other and we have two degenerate sets
with ($R_1=R_4$) and $(R_2=R_3)$. One can then define two parameters
\bea
C_- & \equiv & \cos (\delta_D-\gamma)=\frac{1}{2 r_B r_D}(R_4-r_B^2- r_D^2)\;,
\nn\\
C_+ &\equiv & \cos (\delta_D+\gamma)=\frac{1}{2 r_B r_D}(R_2-r_B^2- r_D^2)\;,
\eea
where, we have deliberately retained the $r_B$ term in the above expressions,
so that one can still use this method for $r_B \neq 1 $ case.
Thus one can now obtain the solution for $\gamma $, in terms of these
observables, as
\be
\sin^2 \gamma= \frac{1}{2} \big [
1-C_+ C_- \pm \sqrt{(1-C_+^2)(1-C_-^2)} \big ]\;,
\ee
one solution of which will give $\sin^2 \gamma $ while the other
being $ \sin^2 \delta_D $. Since $\delta_D$ has already been measured,
$\sin^2 \gamma$ could be extracted from these observables, once we know the
values of $R_2 $, $R_4$ (otherwise $R_1$ and $R_3$) and $r_B$ (it may be
noted that the value of $r_D$ is already known now).

Our method consists of two parts, the first one being the
$B_c^\pm\to D^0(\bar D^0)D_s^\pm$,
which will be measured at the hadron colliders,
such as LHC, whereas the second part
consists of the measurement of $D^0(\bar D^0)\to K^{*+} K^-$, which can also
be measured at the same collider experiments. Moreover, since we have already
experiments and there are upcoming dedicated experiments to measure the
parameters in the charm-sector, like at CLEO-c and the BEPCII, which
will provide us half of the parameters needed in our study, it
is meaningful to combine the data from various experiments, mentioned above,
to obtain $\gamma$ with
a better accuracy. In principle, one can study the $D^0 \to K^+ \pi^0 K^-$
(where $K^{*+}$
decays to $K^+ \pi^0$) but since CLEO and other charm experiments are doing
precisely the same job we, therefore, leave it to these experiments
to provide us the values of $r_D$ and $\delta_D$.

We would like to comment here that the possible effect of $D^0-
\bar D^0$  mixing for the determination of $\gamma$ is not taken into
account in our analysis since it has been well studied in the
literature \cite{mec1, ggsz}
and found that the effect is very small, unless we are doing a precision
measurement of $\gamma$. To be quantitative the error could be around
1$^\circ$, with the present data available, which for all practical
purposes can be ignored at this moment.

Now, with $r_D$ already known (so also $\delta_D$), we are left with only
two unknowns ($\delta_B$ and $\gamma$). Therefore,
we have two unknowns and four observables.
We can consider different non-CP eigenstates (like $\rho^+ \pi^-$),
which will increase the observables by four and
unknowns by two ($r_D^\prime$ and $\delta_D^\prime$) for each additional
eigenstate. One can also take
$B_c^\pm\to D^{0}(\bar D^0)D_s^{*\pm}$ mode thereby further increasing number
of observables by four and unknowns by two (say $r_B^\prime$ and
$\delta_B^\prime$, in fact it could be just $\delta_B^\prime$).
Hence we hope to have
enough observables and at best half the number of unknowns (actually,
it will always be less than half since new unknown parameters, namely,
$r_D^\prime$ and $\delta_D^\prime$ can also be inferred from
the D decay data) and we can obtain the value of $\gamma$
without hadronic uncertainties. Also, it should be reminded here
that by the time the actual measurement could be done, using this method,
results from the other methods, mentioned earlier, might be available.

Now let us estimate the branching ratios for these modes. Using the
generalized factorization approximation, the amplitudes are given as
\bea
{\cal A}(B_c^- \to D^0 D_s^-) &= & \frac{G_F}{\sqrt{2}} V_{cb}V_{us}^*
(a_2^{eff} X + a_1^{eff} Y )\;,\nn\\
{\cal A}(B_c^- \to \bar D^0 D_s^-) &= & \frac{G_F}{\sqrt{2}} V_{ub}V_{cs}^*
(a_1^{eff} X_1 + a_2^{eff} X ),\;,
\eea
where, $X=i f_{D^0}(m_{B_c}^2-m_{D_s}^2)F_0^{B_c  D_s}(m_{D^0}^2)$,
 $X_1=i f_{D_s}(m_{B_c}^2-m_{D^0}^2)F_0^{B_c D^0}(m_{D_s}^2)$and
$Y=i f_{B_c}(m_{D_s}^2-m_{D^0}^2)F_0^{D_s  D^0}(m_{B_c}^2)$ are
the factorized hadronic matrix elements. For numerical evaluation we
use the values of the form factors at zero recoil
from \cite{du} as $F_0^{B_c D^0}(0)=0.352$,  $F_0^{B_c D_s}(0)=0.37$,
the decay constants (in MeV) as $f_{D^0}=235$,  $f_{D_s}=294$,
$f_{B_c}=360$,  the QCD coefficients $a_1^{eff}=1.01$, $a_2^{eff}=0.23$,
particle masses, lifetime of of $B_c$ and CKM matrix elements from \cite{pdg}.
We thus obtain the branching ratios as
\bea
BR(B_c^- \to D^0 D_s^-)=7.0 \times 10^{-6}\;,~~~
BR(B_c^- \to \bar D^0 D_s^-)=4.5 \times 10^{-5}\;.
\eea
Let us now make a crude estimate of the number  of reconstructed
events that could be observable at LHC per year of running. At LHC,
one expects about $10^{10}$ untriggered $B_c$'s per year \cite{bc1}.
For the estimation we use the branching ratios 
as $BR(B_c^- \to D^0 D_s^-)=7.0 \times 10^{-6}$ and
$BR(D^0  \to K^{*+}K^-)=3.7 \times 10^{-3}$ \cite{pdg} and
assume that the $D_s$ can be reconstructed efficiently by combining
a number of hadronic decay modes. As the LHCb trigger system
has a good performance for hadronic modes, we assume an
overall efficiency of $30 \%$ and
hence  we expect to get nearly 80 events per year of running at LHC.

We have outlined here that $B_c^\pm \to (D^0) D_s^\pm \to
(K^{*\pm}K^\mp) D_s^\pm $ and $B_c^\pm \to  (\bar D^0) D_s^\pm \to
(K^{*\pm}K^\mp) D_s^\pm $ can be used to determine the 
CKM angle $\gamma$ at the
LHC. Since the interfering amplitudes are of equal order (which is not the case
with $B\to DK$ methods) and furthermore neither tagging nor
time dependent studies are required to undertake this strategy
and above all the final particles are charged ones (and of course with
reduced background) this method may
be very well suited for the determination of $\gamma$ without hadronic
uncertainties. But one has to pay the price for all the niceties of this
method in the sense that the branching ratios are smaller by an order
compared to the earlier modes. Nevertheless,
we hope that this should not cause any hindrance for the clean determination
of angle $\gamma$ using this method, and even if we get lesser number of events
the predictive power will not be diluted.

In conclusion, in this paper we have looked into the possibility of
extracting the CKM
angle $\gamma$ using multibody $B_c$ decays and in view of the fact that
LHC is coming into operation shortly
this method can be found to be very useful to obtain $\gamma$ in yet
another method
to supplement the results from other methods. We believe that during the first
few years of LHC run we will have a meaningful value of angle $\gamma$ with
reduced errors and
emphasize that the strategy presented here will be an added asset to
our endeavour to measure the angle $\gamma$.

\acknowledgments

The work of RM was partly supported by Department of Science and Technology,
Government of India,
through grant No.
SR/S2/HEP-04/2005. BM would like to thank Council of Scientific
and Industrial Research, Government of India, for financial support.


\end{document}